\documentclass{emulateapj}
\usepackage{natbib}
\usepackage{epsfig}
\begin{document}


\title{Nitrogen Production in Starburst Galaxies Detected by {\sl GALEX}}

\author{ Ryan P. Mallery\altaffilmark{1}, 
Lisa Kewley\altaffilmark{2},
 R. Michael Rich\altaffilmark{1}, 
Samir Salim\altaffilmark{1}, 
Stephane Charlot\altaffilmark{3,4},
 Christy Tremonti\altaffilmark{5},
  Mark Seibert\altaffilmark{6}, 
Todd Small\altaffilmark{6}, 
Ted Wyder\altaffilmark{6}, 
Tom A. Barlow\altaffilmark{6},
Karl Forster\altaffilmark{6},
Peter G. Friedman\altaffilmark{6},
D. Christopher Martin\altaffilmark{6},
Patrick Morrissey\altaffilmark{6},
Susan G. Neff\altaffilmark{7},
David Schiminovich\altaffilmark{8},
Luciana Bianchi\altaffilmark{9},
Jose Donas\altaffilmark{11},
Timothy M. Heckman\altaffilmark{12},
Young-Wook Lee\altaffilmark{10},
Barry F. Madore\altaffilmark{14},
Bruno Milliard\altaffilmark{11},
Alex S. Szalay\altaffilmark{12},
Barry Y. Welsh\altaffilmark{13}, 
Suk Young Yi\altaffilmark{10}}

\altaffiltext{1}{Department of Physics and Astronomy, University of California, Los Angeles, CA 90095-1562}
\altaffiltext{2}{Institute for Astronomy, University of Hawaii, Honolulu, HI, 4234092309239805} 
\altaffiltext{3}{Max-Planck Institut f\"ur Astrophysik, D-85748 Garching, Germany}
\altaffiltext{4}{Institut d'Astrophysique de Paris, CNRS, 98 bis boulevard Arago, F-75014 Paris, France}
\altaffiltext{5}{Steward Observatory, 933 North Cherry Avenue, Tucson, AZ 85721}
\altaffiltext{6}{California Institute of Technology,MC 405-47, 1200 East California Boulevard, Pasadena, CA 91125}

\altaffiltext{7}{Laboratory for Astronomy and Solar Physics, NASA Goddard
Space Flight Center, Greenbelt, MD 20771}

\altaffiltext{8}{Department of Astronomy, Columbia University, New York, NY 10027}

\altaffiltext{9}{Center for Astrophysical Sciences, The Johns Hopkins
University, 3400 N. Charles St., Baltimore, MD 21218}

\altaffiltext{10}{Center for Space Astrophysics, Yonsei University, Seoul
120-749, Korea}

\altaffiltext{11}{Laboratoire d'Astrophysique de Marseille, BP 8, Traverse
du Siphon, 13376 Marseille Cedex 12, France}

\altaffiltext{12}{Department of Physics and Astronomy, The Johns Hopkins
University, Homewood Campus, Baltimore, MD 21218}

\altaffiltext{13}{Space Sciences Laboratory, University of California at
Berkeley, 601 Campbell Hall, Berkeley, CA 94720}

\altaffiltext{14}{Observatories of the Carnegie Institution of Washington,
813 Santa Barbara St., Pasadena, CA 91101}

\begin{abstract}
We investigate the production of nitrogen in star forming galaxies with ultraviolet 
(UV) radiation detected by the Galaxy Evolution Explorer Satellite ({\sl GALEX}). We use a 
sample of 8,745 {\sl GALEX} emission line galaxies matched to the Sloan Digital Sky 
Survey (SDSS) spectroscopic sample. We derive both gas-phase 
oxygen and nitrogen abundances for the sample, and apply stellar population synthesis models 
to derive stellar masses and star formation histories of the galaxies.
 We compare oxygen abundances derived using three different diagnostics. 
 We derive the specific star formation rates 
 of the galaxies by modeling the $7-$band {\sl GALEX}$+$SDSS photometry.
We find that galaxies that have log SFR/M$_*\gtrsim-10.0$ typically have values of log N/O $\sim0.05$ dex less than 
galaxies with  log SFR/M$_* \lesssim-10.0$ and similar oxygen abundances. 
\end{abstract}

\section{Introduction}
           The abundance of nitrogen in
           galaxies and the site of its creation is critical
           for our understanding of galaxy chemical evolution.
           The ratio of N/O is especially useful because both of these elements are
           created by different mechanisms in different ranges of stellar mass.
           Nitrogen is produced during hydrogen burning via the CNO
           and CN cycles, and is created as both a primary and secondary element.
           In primary nucleosynthesis the production of nitrogen is
           independent of the initial metallicity of the
           star. Primary
           production of nitrogen occurs predominantly only
           in intermediate mass stars
           (4$\leq$M$/$M$_{\odot}$ $\leq$8)
           \citep{mat85, mat86} 
           where the nitrogen is produced during hydrogen
           shell burning while carbon and oxygen, which are
           assumed to be primary nucleosynthesis elements, are moved from
           the core to the outer stellar layers during
           dredge-up episodes. Recent stellar models that include
           rotational effects indicate that massive
           stars between 9 and 20 M$_{\odot}$ may produce primary nitrogen \citep{maeder}. These massive star models
           incorporate a convective helium burning shell that penetrates into the hydrogen
           burning shell, creating primary
           nitrogen. The discovery of metal-poor halo stars with high N/O ratios by \citet{spite} and \citet{israelian}
           seems to confirm the primary production of nitrogen in massive stars with a yield that depends on stellar mass and metallicity \citep{chia05, chia06}. 
           In secondary production, nitrogen is synthesized
           from the carbon and oxygen initially present in the
           star and its abundance is therefore proportional to
           the initial heavy element abundance.  Secondary production is common to
           stars of all masses \citep{mat85,mat86}. Though the synthesis of nitrogen in stars is becoming better understood, 
           our understanding of the abundance of 
           nitrogen in galaxies is lacking. One way to investigate  the primary versus secondary origin
           of nitrogen is to examine the ratio of N/O as a function of O/H. These abundance ratios are commonly computed from 
           optical nebular emission lines of HII regions (see \S 5). In the case of primary nucleosynthesis N/O will be constant; secondary enrichment 
           produces a linear correlation between log N/O and log O/H. The combination of primary and secondary nucleosynthesis produces a 
           non-linear relation (see Fig. 3). The abundance ratios of many other elements such as neon, sulfur, and iron  with respect to oxygen have been found to 
          tightly correlate with O/H \citep{iz05}, yet at fixed O/H galaxies have been found to have a scatter in N/O of a factor of 2 \citep{villa93}. 
          Until recently, this problem could not be 
           adequately addressed due to small sample sizes, and uncertain abundances. 
                     
           Chemical evolution scenarios proposed to explain this scatter include (1) a primary plus
           secondary origin of nitrogen but with variable
           initial mass functions (IMFs) \citep{alloin},
           (2) a primary plus secondary origin but with a time delay
           between the release of nitrogen and the release of oxygen back into
           the interstellar medium (ISM) \citep{villa93, garnet90, thurston}. With
           regards to the former scenario,
           \citet{chia} have found that an IMF constant in
           space and time better reproduces the observational
           constraints of the solar neighborhood  (i.e.
           the ratio of metal-poor to metal-rich stars, the ratio of SN II to SN I,
           and the ratio of He to metal abundance.). \citet{chia} find that such an IMF also helps reproduce the observed abundance
           gradient of the Galactic disk more reliably than models with IMFs that depend on metallicity or SFR. 
            In the time delay chemical evolution model,
           oxygen is released in the supernovae
           of short lived massive stars. As the metallicity of successive generations of  massive stars increases,
           secondary nitrogen is also released, and then the bulk of nitrogen is
           released much later in intermediate mass stars. The chemical evolution models of \citet{field}  
            for blue compact galaxies and \citet{chia03} for dwarf galaxies, 
           incorporating the variation of stellar yields with stellar mass and stellar evolution timescales, 
           have shown that the scatter in N/O could be reproduced by varying star formation histories.

           Work by \citet{considere} on
           abundances in barred spiral
           galaxies indicates that the nitrogen abundance
           is the result of both primary and secondary
           nucleosynthesis. However, nitrogen abundances taken from a small UV selected
           galaxy sample detected by FOCA\footnote{FOCA was a
           balloon-borne 40 cm telescope that imaged at
           2015\AA, FWHM 188\AA~\citep{mil}.} \citep{contini} show mostly a secondary component, 
           but still with considerable scatter. They proposed that the difference between the two results arises because there is a time delay
           between the release of oxygen and nitrogen. The UV galaxies were starbursts;
          consequently, the high mass stars formed during the starburst had released newly synthesized oxygen into the ISM.
           This increased the oxygen abundance 
           and lowered the N/O  ratio of these galaxies such that their new nitrogen and oxygen 
           abundances were consistent with only a secondary nitrogen component. Then later, 
              once the intermediate mass stars formed during the burst have had time to
           evolve, nitrogen will be released into the ISM. This would increase the nitrogen in these UV galaxies to a similar  amount of N/O
           as the sample of \citet{considere}.
         
             The time delay scenario, as stated  by \citet{contini}, may be oversimplified. It does not account for 
             galaxies that have  star formation histories differing from cycles of bursting  phases followed by 
             quiescent phases, and was made prior to any evidence for the production of primary 
           nitrogen in massive stars \citep{maeder, spite, chia05,chia06}. A more accurate statement is: the release of material by 
           a star of mass, M into the ISM will cause the N/O ratio of the ISM  to increase if its relative yield of p$_N$/p$_O$ is greater than 1 and decrease
           if its relative yield  is less than 1. The new nitrogen and oxygen stellar yields of \citet{chia06}  
           still indicate that while massive stars (M$_* >9$M$_{\odot}$) produce primary nitrogen, the ratio of nitrogen to oxygen yields for massive stars
            is $\lesssim 1$ and still decreases with increasing mass. Thus, a burst of  star formation will 
            still initially cause a decrease in N/O, 
            the effect of the new yields only diminishes the extent to which a starburst can conceivably lower N/O. 
               All that is required for the N/O  of a galaxy to increase 
            is that its current SFR is less than its past average SFR such that comparatively fewer high mass stars are being formed, allowing 
            for intermediate mass stars of previous generations to 
            dominate the chemical enrichment of the galaxy.

           The large amount of uniformly calibrated data from the Sloan Digital Sky Survey (SDSS) has recently enabled
           robust statistical studies of chemical  enrichment \citep{trem05}.
           Recently, \citet{iz05} examined the ratio of N/O  in metal-poor galaxies, 
           $12+\log O/H <8.5$, in SDSS  DR3 and
           concluded that the N/O ratio increases with increasing starburst age (decreasing EW$_{H\beta}$) 
           for metal-poor galaxies, due to the ejection of 
           nitrogen by Wolf-Rayet stars. \citet{liang}
           consider a  $\sim 30,000$ galaxy SDSS DR2 sample, and show that objects with higher N/O tend to have lower $EW_{H\alpha}$.
        This result is consistent with  those galaxies 
            with current star formation rates that are high with respect to their past average star formation rate, exhibiting a higher
           oxygen abundance. The oxygen is presumably contributed by the recently formed massive stars.
            Here we examine the N/O ratio as a function of
            O/H in a sample of UV selected galaxies
           detected by the Galaxy Evolution Explorer (GALEX),
           exploiting the overlap between the {\sl GALEX} Medium Imaging Survey and the SDSS
           spectroscopic footprint in the local
           universe ($z<0.3$). The large volume of data available from {\sl GALEX}+SDSS
          makes it possible to go beyond equivalent widths and calculate physical quantities like stellar masses, M$_*$, and specific
          star formation rates, SFR/M$_*$, which can be more easily compared with models when investigating chemical enrichment.
              We use the SFR/M$_*$ and  $M_*$, of these galaxies
           derived from their 7-band UV-optical photometry \citep{ss07} to test whether the star formation history
           of a galaxy can explain the
           observed relationship between the nitrogen and oxygen abundances in our sample and 
            investigate the accuracy of determined abundances.  
         We note that O/H is not equivalent to a time axis, and the values of  N/O and  O/H 
         represent the current chemical evolutionary stages for galaxies that 
         have most likely had different histories of star formation and other dynamical processes such as galactic winds
         and gas accretion timescales \citep{chia03, diaz86, mat85}. We expect, nevertheless, that if the time delay scenario
         is correct, that galaxies currently exhibiting a 
         strong burst of star formation will on average have lower values of N/O than non-bursting galaxies at similar metallicities.

          The outline of the paper is as follows. In
          \S 2 we describe the data and sources used in this
          analysis. We present our galaxy
          sample containing matched GALEX and SDSS sources in
          \S 3. An explanation of the derivation of galaxy
          parameters by matching the models of \citet{bc03} to
          the 7 color UV-optical SED of each source is
          given in \S 4. We describe and contrast the various
          methods used in determining oxygen
          abundances in \S 5 and nitrogen in \S 6. In
          \S 7  we examine the relationship of nitrogen to oxygen for our sample  and
          in \S 8 we give our conclusions. We assume {\it H$_o$}=70
          km s$^{-1}$ Mpc$^{-1}$, $\Omega_{m}$=0.3, and  $\Omega_{\Lambda}$=0.7.

\section{Data}
   We consider galaxies with GALEX photometry from the Medium
  Imaging Survey (MIS) Internal Release 1.1 (IR1.1), m$_{lim}$(AB)$\approx$ 23, and SDSS
  photometry and spectra.  {\sl GALEX}  is a NASA Small
  Explorer Mission that aims to survey the UV emission from Galactic and extragalactic sources
  from a 700km circular orbit
  \citep{martin,pm}. {\sl GALEX}  images the sky simultaneously in two bands, 
  the far-UV (FUV 1344-1786\AA) and the near-UV (NUV 1771-2831
  \AA). Each {\sl GALEX}  circular field is 1.25 deg. in  diameter. We use
  FUV and NUV magnitudes and magnitude errors
  derived in elliptical apertures\footnote{{\sl GALEX}  source
  detection and measurement is obtained from SExtractor
  \citep{bar}}. 

   We use optical photometry for our objects obtained from
   SDSS Data Release 4 (DR4) spectroscopic sample. Most of our objects were taken from
   the main galaxy spectroscopic survey ({\it r}$_{lim}<$17.8), but many of our objects were originally 
   targeted as quasars and taken from the quasar spectroscopic survey  ({\it r}$_{lim}<$19.5 \citet{york}). 
   The SDSS photometric data are taken with
  the 2.5m telescope at Apache Point Observatory. Imaging is
  obtained in  {\it ugriz} bands
  \citep{fuk96,smith02}. The imaging data are photometrically
  \citep{hogg} and astrometrically \citep{pier} calibrated. An
  overview of the SDSS data pipelines and products can be
  found in \citet{stoughton}. 
  
 The SDSS spectra are acquired using  3\arcsec~diameter fibers
  positioned on the centers of the target galaxies. The spectra are flux and wavelength calibrated for
  wavelengths between 3800-9200\AA~at resolving power \citep{york}
  R$\equiv$$\lambda$$/$$\Delta$$\lambda$ =1850-2200. 
  We use continuum subtracted emission-line fluxes and flux errors from the
  SDSS spectra measured by \citet{trem05}, to divide and classify the
  sample in terms of emission-line ratios, and to derive
  nebular abundances.  

\section{The Sample}
   We use a sample constructed by matching objects with {\sl GALEX} MIS IR1.1
   detections to galaxies in the SDSS DR4 spectroscopic sample. The objects are matched 
within a 4\arcsec~radius \citep{sbrt05,ss07}. We accept only
unique matches and discard objects that contain multiple matches. 
We restrict the sample to galaxies with z$<0.3$ 
that are detected by {\sl GALEX} in the NUV at a 3$\sigma$ level.
We further restrict the sample to galaxies with spectral {\it r}-band fluxes greater
than  20\% of their total {\it r}-band fluxes. \citet{kjg} found that in samples where
the spectroscopic fiber collects greater than 20\% of the
galaxy light, the fiber metallicities approximate global values. This criterion gives a sample of 36225 objects. 

         In order to constrain the errors on the derived abundances for objects, we impose detection
         criteria for several emission lines. We remove galaxies
         with $<5\sigma$ detections of the Balmer lines H$\alpha$ and H$\beta$
         and [NII]$\lambda$6584.
         For other oxygen forbidden
         lines that are used in the analysis, [OII]$\lambda$3726,3729 [OIII]$\lambda$5007, 
         we remove galaxies with $<3\sigma$ detections. We note that
         demanding a [OII]
         detection restricts the galaxy's redshift
         to $z>0.03$ due to the wavelength cutoff of
         the SDSS spectrograph at 3800\AA. These constraints
         trim the sample to 12213 galaxies.
         
         We identify Active Galactic Nuclei (AGN) in our  sample 
          by using  the line diagnostic diagram
         [NII]$/$H$\alpha$ versus [OIII]$/$H$\beta$ \citep{bpt}.
         We use the formula of
         \citet{kauf03b} to remove galaxies with contributions to their emission line spectrum from AGN.
         The fraction of galaxies removed
         because of possible contamination due to AGN is $\sim$21\%.
         
       Other sources of emission-line flux besides star forming regions include
       planetary nebulae (PN), and supernova remnants (SNR). Studies by \citet{oey} of the Large
       Magellanic Cloud (LMC)  show that SNRs affect the
       emission-line spectra at a fairly low level. As
       discussed in \citet{cl01} the radiation from planetary
       nebulae can be neglected since the ionizing radiation is
       typically less than 0.1 percent of that produced by
       massive stars at an earlier age. 

         We finally remove 831 galaxies with failed fits to the photometric 7-band SED that give reduced $\chi^2 >10$.   
         This gives a sample of 8,745 galaxies, from which 72\% are from the SDSS main galaxy sample.
          Galaxies in the final star forming sample with 3$\sigma$ FUV
         detections comprise $\sim$84\% of the sample.  The emission-line criteria
         we use selects galaxies that are blue in NUV$-{\it
         r}$, with $NUV-r < 4$. {\sl GALEX} is remarkably sensitive to  star-forming galaxies. In all of the {\sl GALEX} MIS 
         IR1.1 fields only 155 galaxies that are detected in SDSS DR4 and that satisfy our emission line criteria do not have
         3$\sigma$ NUV or FUV detections. The percent {\sl GALEX} detection is $99.4$\% for  ${\it r}<17.8$ and $97.9$\% 
         for ${\it r}>17.8$.

\section{Derived Galaxy Parameters}     
   We use the following galaxy parameters derived by
   \citet{ss07}: the NUV and FUV dust attenuations, A$_{NUV}$
   and A$_{FUV}$ in magnitudes, the current star formation
   rate, SFR, averaged over the past 100 Myr in M$_{\odot}$ yr$^{-1}$, the
   present-day stellar mass, M$_*$, of the galaxy in
   M$_{\odot}$, the specific star formation rate, SFR/M$_*$, and the fraction of stellar mass
  formed in starbursts over the last 100Myr, F$_{Burst}$. 
 Galaxy parameters are derived from model libraries of
 galaxies at redshifts of .05, .10, .15, .20, and .25. Each
 library consists of $\sim$10$^5$ models. Each model is
 defined by several parameters: galaxy age, optical depth, star formation history, and metallicity. 
 The star formation history of each model follows the prescription of  \citet{kauf03a} and consists of an underlying, continuous, exponentially
 declining SFR upon which bursts of star formation, random in time and amplitude, are superimposed. 
 Dust attenuation in each model is parametrized using the
 prescription of \citet{cf00}. A description of the prior distributions of
 the model parameters is discussed in \citet{ss05} and \citet{ss07}.
 
  Model spectral energy distributions (SEDs) are created for each set of model parameters
  using the population synthesis code of \citet{bc03} and
  assuming a \citet{kroupa} IMF.    
  The model SEDs are convolved with the GALEX and SDSS filter
  response curves. Statistical estimates of physical galaxy parameters  are derived by
  comparing the observed 7 band GALEX$/$SDSS  fluxes of each
  galaxy to all the convolved model SEDs in the nearest redshift
  library. Probability density functions  (PDFs) for each
  physical parameter are created by assigning weights to
  the parameters of a model. The $\chi^{2}$ goodness of fit of each
  model determines the weight ($\propto \exp[-\chi^{2}/2]$)
  that is assigned to the parameters of that model. The median
  of the PDF is taken as the estimate of the galaxy parameter.
  An estimate of the error for the parameters is taken as
  $1/4$ of the  2.5-97.5 percentile range. Table 1 lists the
  parameters and their mean errors.

 \section{Oxygen Abundance}
       In order to estimate the abundance of oxygen
       we employ three  methods using relations of various
       emission-line fluxes: the R$_{23}$ strong-line abundance
       calibration  of \citet{mcgaugh},
       the O3N2 strong-line calibration of
       \citet{pp04} and the Bayesian metallicity estimates of \citet{trem05}.

        We use the following flux ratios in our calculations dereddened using the extinction curve of Seaton (1979), assuming R$_v=3.1$ and 
        Case B recombination \citep{oster}:
             \begin{equation}
      \frac{[OIII]}{[OII]}\equiv\frac{[OIII]\lambda5007}{[OII]\lambda3727}
              \end{equation}
              \begin{equation}
        O_{32}\equiv\frac{[OIII]\lambda5007
                 +[OIII]\lambda4959}{[OII]\lambda3727}
              \end{equation}
               \begin{equation}
\frac{[OIII]}{H\beta}\equiv\frac{[OIII]\lambda5007}{H\beta}
              \end{equation}
              \begin{equation}
\frac{[NII]}{H\alpha}\equiv\frac{[NII]\lambda6584}{H\alpha}
              \end{equation}
              \begin{equation}
\frac{[NII]}{[OII]}\equiv\frac{[NII]\lambda6584}{[OII]\lambda3727}
              \end{equation}
             \begin{equation}
R_{23}\equiv\frac{[OII]\lambda3727+[OIII]\lambda5007+[OIII]\lambda4959}{H\beta}
              \end{equation}
             \begin{equation}
 O3N2\equiv\log( \frac{[OIII]\lambda5007/H\beta)}{[NII]\lambda6583/H\alpha})
              \end{equation}
               
        The strong line abundance calibration was first
       developed by \citet{pag79} and \citet{alloin}. 
       The various line ratios that have been used to
       calculate abundances are [NII]$/$H$\alpha$,
       [OIII]$/$[NII],
       [NII]$/$[OII], ([SII]
       $\lambda\lambda$66717,6731+ [SIII]
       $\lambda\lambda$9069,9532])$/$H$\beta$, and R$_{23}$,
        which was first introduced by \citet{pag79}. Oxygen strong line
       abundance calibrations are either (1) based on
       photo-ionization models \citep{diaz,kd03} or (2) on abundances
       measured in nearby HII regions where the electron
       temperatures of the ionized regions can be measured. The latter method
       requires detection of faint auroral
       emission lines (e.g.,[OIII]$\lambda$4363, [NII]$\lambda$5755) to determine the electron
       temperature, {\it T$_e$}. The empirical strong-line analytical
       expressions are created from these measurements to
       estimate the abundance in galaxies and HII regions
       that lack significant detections of the auroral lines
       but have similar abundances.

        Recently the studies of
       \citet{kbg,bgk,gkb} using electron temperatures for
       high metallicity HII regions in M101 and M51,
       indicate that various strong-line methods calibrated to photo-ionization models (e.g,
       \citet{mcgaugh}) estimate a higher abundance at high
       metallicities by 0.2-0.5 dex than the {\it $T_e$} method abundances.
       It is currently not clear which method is correct. There is some evidence that the abundances calculated by the {\it $T_e$} method 
     may be underestimated due to temperature fluctuations in the ionized regions causing the electron temperatures to be overestimated,
     and that the strong line abundances may be more correct since their line ratios are not as temperature sensitive as 
     [OIII]4363 \citep{peimbert, bres, bgk,kbg}. 
  
      In our analysis we use two strong-line calibrations to estimate the oxygen abundance: the R$_{23}$ diagnostic of \citet{mcgaugh} calibrated 
      from photo-ionization models,
      and the strong-line ([OIII]/H$\beta$)/([NII]/H$\alpha$) diagnostic of \citet{pp04} calibrated from {\it $T_e$} abundances. 
        The derived abundances for each method are respectively labeled M91, and O3N2 throughout the 
      remainder of this work. We add to these strong line abundances the strong line oxygen abundance estimates of 
       \citet{trem05} (hereafter labeled T04), obtained from
       likelihood distributions of oxygen abundances derived by
       matching emission line fluxes from integrated galaxy spectra
       models of \citet{cl01} to the measured fluxes.

      The R$_{23}$ analytical
      expressions calibrated by \citet{mcgaugh} and given in
      \citet{kobul} that we use to calculate M91 are cited below. Many
      other authors have developed techniques for estimating
      abundances from R$_{23}$. Examples include \citet{cl01},
      \citet{alloin}, and \citet{EnP}. R$_{23}$ is useful
      because it provides an estimate of the total cooling due
      to oxygen. The major caveat  with $R_{23}$ is that it is
      double valued with respect to metallicity. At low oxygen abundances, $12+\log O/H\lesssim8.4$, R$_{23}$
      increases with rising abundance until $12+\log O/H\gtrsim8.4$, after which it begins to
      decrease as metals begin to cause efficient cooling,
      lowering the electron temperature and thus decreasing
      the amount of collisional excitation of the oxygen ions. The metal-poor branch expression is:
      
     \begin{equation} 
         12 + \log(O/H) =7.065+.767x+.602x^2- y(.29
         +.332x-.3318x^2)
             \end{equation}
    and the metal-rich branch expression is:
        \begin{equation}
         12+\log(O/H) = 9.061-.2x-.237x^{2}-.305x^{3}-.0283x^{4}- 
              y(.0047-.0221x -.102x^{2} -.0817x^{3}-.00717x^{4})
         \end{equation}
 
        where x$\equiv\log(R_{23}$) and
        y$\equiv\log(O_{32}$). O$_{32}$ is used to correct
        the effect of the ionization parameter  on R$_{23}$. \citet{kd03}
        have found that the O$_{32}$ ratio depends on
        metallicity and as a result is not a good indicator of ionization
        unless an initial estimate of metallicity can be
        given and an iterative process is applied.   

       To determine on which branch the correct solution
       lies, we use the metallicity sensitive ratios [NII]$/$H$\alpha$ and
       [NII]$/$[OII]. For
       $\log([NII]/H\alpha$) $<$-1 and $log([NII]/$[OII]) $<$ -1.5
       we use the metal-poor expression. For $\log[NII]/H\alpha$ $>$-1 and $\log[NII]/$[OII] $>$ -0.8   
       we use the metal-rich expression. If  -0.8$>\log(
       [NII]/$[OII]) $>$ -1.5, then  we use the  $\log([NII]/H\alpha)$ ratio as stated above to determine the correct branch. 
        Where the two ratios give conflicting
       estimates, the average of the two expressions is used to
       derive the abundance. This is because the solutions for the two branches converge at intermediate metallicities, 
      $12+\log O/H\sim8.4$, and
       it is near this metallicity where the
       metallicity sensitive ratios  are likely to  give a conflicting answer. The average of the two branch solutions in this 
       case should minimize any bias
       in the calculations.
       A caveat with this procedure is that galaxies with intermediate oxygen abundances 
       but with high SFRs will have lower ratios of [NII]/H$\alpha$ and as a result
       can have their abundances calculated with the lower branch and therefore underestimated.

      The strong line calibration O3N2  developed by
      \citet{pp04} is shown below. The calibration based on
      this flux ratio also has several problems. First, it is not corrected for ionization parameter. Second, it is
      based on the flux from a forbidden nitrogen line whose
      abundance many authors claim depends star-formation history of
      the galaxy. As a result the calibration is accurate to
      $\log (O/H) =\pm0.25$, and is only valid for
      O3N2$< 1.9$ (e.g. $12 + \log(O/H) \gtrsim 8.1$). 
       \begin{equation}
  12 +\log(O/H) = 8.73-0.32 \times O3N2      
       \end{equation}
       There is also evidence that at metallicities $\gtrsim
       \log(O/H)_{\odot}$ (O3N2 $\lesssim 0.4)$ the O3N2
       calibration overestimates the
       oxygen abundance \citep{bgk}.            
   
       Figure 1 shows the difference between all  
       the abundance calibrations as a function of stellar mass. 
       In the figure we have transformed each panel into a 75 by 75 pixel image. The mean SFR/M$_*$ of the points in each pixel 
       is shown in true color representation. The mean difference between M91 and O3N2 shows some dependence on stellar mass, with the lower 
          branch of M91 giving lower abundances than O3N2, typically
        about $0.1$ dex with a dispersion of $0.18$ dex. The upper branch of M91 calculates larger abundances than O3N2 typically by $\sim0.2$ dex with a
        dispersion of $0.14$ dex. 
       While the offset between O3N2 and M91 shows little dependence on stellar mass, 
      the offset between T04, and the M91, O3N2 diagnostics show a dependence on stellar mass. As galaxy mass increases, 
      T04 estimates an increasingly larger metallicity than the other two calibrations.
        
\section{Nitrogen Abundance}
    We calculate nitrogen abundance estimates by first using the calibration of \citet{thurston} to estimate the temperature in the
    [NII] emission region using their calibrated empirical relation created from photo-ionization models:
         \begin{equation}
                   t_{[NII]}=0.6065+0.1600x
                   +0.1878x^2+0.2803x^3
            \end{equation}
     
    where $x\equiv\log R_{23}$.
    We then use this
    temperature to determine the the ratio of N$^+$/O$^+$ based
    on the empirical calibration of \citet{pag92} based off of {\it T$_e$} abundances:
               
                  \begin{equation}
                     Log\frac{N^+}{O^+} = Log
                     \frac{[NII]}{[OII]} + .307 - .02Log t_{[NII]} - \frac{0.726}{t_{[NII]}}    
                  \end{equation}
      
      We finally assume that N$/$O=N$^+/$O$^+$. \citet{thurston} found
      through modeling that this assumption is reliable, with only small uncertainties,
      $\sim$.05 dex. \citet{garnet90} concurs that the  N$^+/$O$^+$
      is an accurate N$/$O indicator for low abundances or
      where the ionizing stars are hotter than
      40,000K. Results of modeling by \citet{stasinka} show that even at high abundance, equating the ion
      ratio to the element ratio is good to within 5\%.

       We also calculate the nitrogen and oxygen  abundance via the {\it T$_e$} method
          for the 33 objects in our {\sl GALEX} emission line sample having at least a
       3$\sigma$ detection of [OIII]43643 to ensure reliable estimates of the electron temperature in the [OIII] ionization regions. Table 2 
       shows the derived ${\it T_e}$ abundances and derived galaxy parameters from SED fitting for these objects. We use the TEMDEN
      procedure in the IRAF package Nebular \citep{fivel} to derive the electron temperature from the ratio of ([OIII]5007+[OIII]4959)/[OIII]4363. 
      The electron temperature in the [OII] and [NII] regions were then estimated using the linear relation from Garnett (1992) to convert the mean 
     [OIII] electron temperatures into mean  electron temperatures in the [OII] ionization regions. We then assume that since [NII] and [OII] have relatively
     similar ionization energies that the [NII] electron temperature equals the [OII] electron temperature. The abundance of each ion O$^{2+}$, O${^+}$, 
     and N$^+$ were then calculated using the IONIC procedure in Nebular. 
     
      All four oxygen calculations show a small abundance range for this sub-sample of objects, thus limiting our ability to determine
      if the difference between the two methods has any dependence on abundance or on N/O. 
      The N/O ratio calculated with the strong line calibration
      shows that it is typically $\sim0.1$ dex greater than the ratio determined by the {\it $T_e$} method with a dispersion of $0.07$ dex.  
      How accurate the strong line N/O ratio is for higher oxygen abundances is unknown, and its precision is
      lacking. The mean error on log N/O for the entire sample is 0.17 dex, due mostly  to the error on the R$_{23}$ temperature. 
      For the the purposes of the rest of our analysis, the accuracy of the strong line nitrogen diagnostic does not matter,
      only the relative difference between each galaxy is of importance.

\section{The N/O versus O/H Relationship}                     
            The time delay scenario for the production of nitrogen predicts that starbursting galaxies exhibit a rise in oxygen abundance
          along with a drop in N/O \citep{contini, vzee06, henry99}. The addition of UV data from {\sl GALEX} 
          to the 5 band SDSS photometry makes it possible to distinguish between galaxies recently hosting
          starbursts and those with declining star formation, because the FUV passband is responsive to star formation
          on timescales of 10 Myr and the NUV passband on timescales of 100 Myr \citep{martin, bc03}. We use our O/H and N/O estimates along with the 
          results from the Bayesian broad band SED analysis 
          to examine if the relative abundance of nitrogen to oxygen in a galaxy can be explained by the galaxy's star formation history. 
                      
           The relationship between N/O and O/H for our sample is shown
          in Figure 2. The points have been pixelated and then scaled by color
          to show the mean value of specific star formation rate
          of the points in each pixel. The specific star formation rate indicates the relative number of recently formed ($\sim100Myr$) high mass stars
          to the cumulative number of stars formed over a galaxy's star formation history. 
          Galaxies with large specific star formation rates have recently undergone a burst of star formation 
          or have a slowly declining SFR. The mean standard deviation in  SFR/$M_*$ for each pixel is $\sim0.07$ dex.
           In Figure 3 we plot our nitrogen-oxygen relationship again
          with the data points plotted as a shaded 2D histogram to aid the interpretation of the previous figure.

           In the figures we  have included the simple closed box 
          model of \citet{villa93} for the primary (solid
          line), secondary (dashed line), and primary +
          secondary (dashed-dotted line) production of
          nitrogen. This model assumes that nitrogen has both a
          primary and secondary component, and that oxygen has
          only a primary component.  The time rate of
          change of each element is taken to be proportional
          to the star formation rate which is assumed to equal a constant times the fraction
          of galaxy's mass in gas ($=1$ at t$=0$).
           Assuming that there are 
          no time delays in the release of the material, a solution for the model can be 
         found,  $\log [N/O] = \log[a+b \times [O/H]]$, 
          where a is the primary yield of nitrogen divided by
          the yield of oxygen and b is the secondary yield of
          nitrogen divided by the yield of oxygen. \citet{villa93} quote
          values of a=.034 and b=120 using a  by-eye fit
          to line strengths taken from literature  for HII
          regions in nearby galaxies. 
          
          The three oxygen abundance methods, allowing for the relative offsets between each method,
           are all consistent with galaxies containing primary nitrogen 
           at low metallicities and a secondary component at higher metallicities. The mean scatter of N/O as a
           function of oxygen abundance is 0.08 dex for O3N2, 0.11 dex for M91 and 0.13 dex for T04.
          We note that these three derivations for oxygen abundance are not completely independent of the nitrogen abundance.
          The O3N2 value depends on a flux ratio containing [NII]6584, while $R_{23}$ is used to calculate 
          the [NII] temperature required to determine the N/O ratio. Furthermore, 
          the models used to derive T04 have prior distributions of metallicities where nitrogen abundance is selected to have a only a primary dependence on
          the oxygen abundance below $12+\log$ O/H$<8.25$, and a completely secondary dependence for metallicities greater than this.
           T04 and O3N2 diagnostics are therefore predetermined to exhibit secondary nitrogen production, 
           and are not useful in determining
          the relative amount of secondary or primary nitrogen in each galaxy. Furthermore, the abundances determined by T04  may
          slightly overestimate abundances for galaxies that have an increased N/O ratio from primary+secondary nitrogen. The likelihood estimates of T04
          depend on the flux from nitrogen emission lines,  but only consider that nitrogen is secondary in origin, and do not accurately account for
          nitrogen fluxes from galaxies containing secondary plus primary nitrogen. Of the three diagnostics, the M91 calculation 
           has the least dependence 
           on the N/O diagnostic. This is because the N/O calculation depends slightly on the temperature estimate obtained by $R_{23}$, which introduces
         a scatter in N/O that increases from $\sim0.04$ dex at the lowest values of N/O to $\sim0.1$ dex at highest N/O values. The
          fact that O3N2 and T04 show a similar secondary dependence on N/O as M91 is an indication that the interdependence between the N/O 
          calculation and T04 or O3N2 is only a small effect.

\subsection{N/O and SFR/M$_*$}

         The main results emerge when we consider the relationship between N/O and O/H as a function of specific star formation rate.
          Figure 2 gives several interesting results. First, for galaxies with high abundances ($12+\log$ O/H$\gtrsim8.6$),
          the M91 and T04  diagnostics both indicate that for galaxies with similar O/H, the most extreme starbursts (highest values of SFR/$M_*$) tend to have
          lower N/O. This is shown more clearly in Figures 3 and 4. In these figures we divide our emission sample into sub-samples of 
          specific star formation rate:   
           log SFR/M$_* < -10.1$ (red points),  $-10.1<$log SFR/M$_* <-9.1$ (green points), and log SFR/M$_* > -9.1$ (blue points). We then calculate the mean 
         N/O value and error on the mean for each of the sub-samples in increments of $0.1$ dex in $\log$ O/H. 
         Figure 3 plots the mean N/O values for each sub-sample 
         as a function  of metallicity, and figure 4 plots the difference between the mean N/O value in each sub-sample  
          and the mean N/O value for the entire sample. Tables 3, 4 and 5 
         list the mean values of N/O, the errors on the mean,
         and the number of galaxies for each bin of O/H with more than 30 galaxies in each sub-sample. All three diagnostics show that the galaxies
         with  log SFR/M$_* > -10.1$ have lower N/O values than galaxies in the lowest specific star formation rate sub-sample (log SFR/M$_* < -10.1$)
         in each decrement of metallicity between $8.5$ and $9.0$ dex. As metallicity
         increases, and nitrogen becomes largely secondary in origin, and 
         the difference between the N/O ratios of the sub-sample with the lowest specific star formation rates and the other two sub-samples  decreases.
                    
          The M91 and T04 diagnostics also show that the most extreme starbursts (log SFR/M$_* > -9.1$) 
          at intermediate metallicities on average have N/O ratios $0.02$ dex lower than galaxies with average specific 
         star formation rates, $-10.1<$log SFR/M$_* <-9.1$, a decrease in N/O of $\sim3\%$.
          The O3N2 diagnostic shows the opposite trend of the other two diagnostics since O3N2
          is not corrected for ionization parameter. 
          At the lowest and highest metallicities
          no difference is found between the N/O ratios of the two sub-samples in all three diagnostics.
 

            Our findings are consistent with  similar conclusions reached by \citet{contini}.            
         The galaxies with the lower specific star formation rates
           have the highest N/O ratios because they  are currently forming comparatively fewer
          high mass stars. This allows the chemical enrichment of the galaxy to be dominated by the products of intermediate mass stars, 
          which generate  more nitrogen than oxygen, causing N/O in these galaxies to rise.
           At low metallicities no difference is found between 
           the N/O ratios of the  the most extreme starbursts and the average star-forming sub-samples. This is because 
           there are only a small number of galaxies in our sample with low metallicities
           and most of these are in the sub-sample with the highest specific star formation rates. At high metallicities, no difference is found  between the 
           the most extreme starbursts and the average star-forming sub-sample. This is presumably because the oxygen generated by  
           the high mass stars formed during the latest starburst constitutes only a small fraction of the total oxygen abundance of the galaxy 
           and has little effect on lowering the N/O ratio. At intermediate metallicities the oxygen abundance is relatively small,
           such that the oxygen created in a starburst
           constitutes  a large fraction of the oxygen abundance of a galaxy and causes a larger decrease in the N/O ratio.  
       
         Other possible explanations for the N/O ratios are variable IMFs and galactic winds. 
         An IMF that produces more massive stars for  galaxies  with higher  specific star formation rates, could possibly cause the low N/O ratios
       of the strongest  starbursts  seen in Figures 2,3, and 4. 
          Such a variable IMF could plausibly have a slope parametrized by either metallicity, SFR, or both. 
          \citet{silk} conjectures that the IMFs of starbursts may be weighted to
          form more massive stars, and several authors have previously parametrized IMFs with a dependence  on metallicity \citep{matt_t, scully}. 
          At this time, the validity of a variable IMF and its effect on the abundances cannot be assessed.
           We find no need to invoke a variable  IMF to model the UV and optical SEDs  of these galaxies. Furthermore,
           \citet{chia} have found that that chemical 
         evolution models for the Galaxy that use a metallicity dependent IMF
          do not adequately reproduce the observational constraints of the solar neighborhood.

         Galaxies with high specific star formation rates could also have galactic winds that differentially remove one element with respect to the other.
         With regards to differential flow of oxygen, 
         \citet{vzee06} examined the ratio of N/O for a sample of dwarf galaxies, and argues that their data suggests that either differential outflow 
         of oxygen occurs in every  galaxy in their  sample
         with the same efficiency or that differential outflow of oxygen has a negligible effect on N/O ratios.  
         They found that the correlation of oxygen abundance with optical luminosity for their sample had a lower 
         scatter than the correlation of nitrogen abundance with optical luminosity. They argue that if differential outflow was the cause of the scatter 
         in N/O, then the oxygen-luminosity correlation should have a larger scatter than the nitrogen-luminosity correlation, since the
         the outflow of oxygen would depend on other galaxy parameters such as galaxy mass, and ISM structure. Nitrogen may be differentially removed in 
         galaxies that have high specific star formation rates, but there is no reasonable explanation as to why this might occur. 
         In fact, one would expect the opposite, that  
         oxygen and not nitrogen  would be differentially removed in starbursts since the kinetic energy responsible for ejecting the material likely comes from
          the supernovae of high mass stars that produce very little nitrogen with respect to oxygen.   
        
\subsection{N/O vs M$_*$ and {\it g-r} optical color}           
           In order to further test the above explanation, we plot the nitrogen to oxygen  relationship again,
           with the pixels scaled by color with mean values of stellar mass, and {\it g-r} color in Figures 5 and 6 respectively.  \citet{trem05}
          found a tight correlation of $0.1$ dex between increasing stellar mass and oxygen abundance, so we expect that mass will increase
           with O/H. Figure 5 shows that the stellar mass increases with O/H with little dependence on N/O for all three methods. 
          Along the same lines, \citet{vzee06} analyze a dwarf galaxy sample and conclude that a trend of increasing N/O 
             correlates with redder B-V color and hence, lower star formation rate.  
           Based on this result, we expect that increasing values of N/O should correlate with redder g-r color.  
            In figure 6 we see that as N/O increases 
          the average {\it g-r} color increases for T04, M91, and O3N2 at high metallicities. At metallicities below $12+\log$ O/H$\sim8.4$ both 
          M91, and T04 show this
          trend. O3N2 does not because it is not corrected for ionization effects. All methods of determining the oxygen abundance show a dependence
          on mass irrespective of their nitrogen abundance. The T04 and M91 diagnostics 
          show that galaxies of similar metallicity but with higher N/O values
          have redder {\it g-r} color. These results suggest that the trend between higher specific star formation rates and lower N/O values is a real trend, 
          but more reliable and consistent metallicity diagnostics are required to test this result.

\subsection{{ \it T$_e$} sample  N/O vs O/H}          
         As a further check, we use the ${\it T_e}$ method abundances to determine whether the trend 
         for galaxies with higher SFR/M$_*$ to have lower N/O ratios is genuine, and not 
         produced by the strong line abundance calculations themselves due to a dependence on an unknown  galaxy parameter.
         The abundances derived from the {\it $T_e$} method are dependent only on electron
         temperature and density, and the N/O ratio calculated by this method is not predetermined to show secondary dependence.   
          In Figure 7 we show the N/O ratio versus  O/H for the 33 galaxies with abundances measured
          by the {\it $T_e$} method. In the upper left the points are colored by their {\it g-r} optical colors.
          Even though there is a good deal of scatter in the figure, the galaxies that
          have the lowest N/O ratios tend to be the bluest in {\it g-r}, with the mean value of ${\it g-r} = 0.12$ and 
          a standard error on the mean of $.04$ for galaxies with 
         $\log N/O < -1.5$ and  {\it g-r} $= 0.2$ with a standard error on the mean of $0.04$ for galaxies with $\log N/O >-1.5$.  These galaxies also tend to have
          slightly higher specific star formation rates  with a mean difference of $0.12$ dex between galaxies with  
         $\log N/O < -1.5$  and galaxies with  
         $\log N/O > -1.5$.  The galaxies with lower N/O ratios also have slightly higher H$\alpha$ equivalent widths, which is an indicator of the current 
          star formation relative to past star formation, on timescales of 10 Myr. 
           The SED fitting indicates that to a 95\% reliability at least half 
          of these galaxies formed 1\% (and as much as $\sim50\%$) of their stellar mass in bursts within the last 100Myr.
           If there is indeed a time delay between the release of oxygen from the massive stars 
          and nitrogen from the intermediate mass stars, then these starburst galaxies should have an influx of newly synthesized oxygen that 
          will raise the oxygen abundance and reduce the N/O ratio.  The results from the {\it $T_e$} abundances
            slightly favor this scenario, but due to the small 
          sample of galaxies with 3$\sigma$ detections of [OIII]4363, the small range of specific star formation rates,
           and the uncertainties on the abundances, we are unable to discern if the star formation history is really the cause 
          of the scatter of N/O values  for the {\it $T_e$} sample. 
         We would expect that since all of the 33 galaxies are large starbursts, that the N/O ratios would lie close to the secondary
          nitrogen curve. The explanation posited by \citet{iz05} for the scatter in N/O for galaxies with similar metallicities 
          is that the Wolf Rayet stars in these galaxies have released a significant amount of nitrogen from winds, which 
          being an order of magnitude more dense than the surrounding ISM  can cause the N/O ratio to appear high, increasing by as much as $0.23$ dex.
           The N/O ratio will decrease as the nitrogen from the WR winds has time to diffuse
          into density equilibrium with the ISM, raising the overall ISM log N/O ratio by $0.03$ dex. However, the chemical evolution models of \citet{chia03} show 
          that the scatter in N/O ratios of dwarf galaxies at metallicities similar to those in our  
           our {\it $T_e$} sample can be explained by different star formation histories, 
           different burst strengths, and burst durations; they show that there is no need to invoke nitrogen from winds of massive stars to show this effect.
          The galaxies with $\log N/O >-1.5$ have on average
           only slightly  bluer optical colors than the rest of the {\it $T_e$} galaxies. This indicates that intermediate mass
         stars from the last major star formation event may be responsible for the high N/O ratios, but does not rule out that winds could cause 
         a portion of the observed scatter, (though the errors on the measured abundances 
          are able to account for a large portion of the observed scatter in N/O). To resolve this issue, and effect of variable IMFs or 
          other dynamical processes on N/O such as
          mixing timescales of the newly synthesized material, 
          more reliable and consistent metallicity measurements             
          are needed,  with errors in the derived abundance less than
          $0.1$ dex. New nitrogen diagnostics 
          for metal rich galaxies are particularly needed to compare our results obtained using the \citet{pag92} strong line diagnostic.
         These results should also be compared to chemical evolution models to substantiate their validity.

\section{Conclusion}
   We consider galaxies detected by
   {\sl GALEX} in the Medium Imaging Survey to a limiting magnitude of $NUV=23 (AB)$. We match our UV star-forming galaxies
   to ${\it z}=0.3$ with the SDSS DR4 spectroscopic sample. \\ \\

    1. {\sl GALEX} at MIS depth ($NUV_{limit}\sim23.0 AB$) 
       detects 98.4\% of star forming SDSS galaxies in the DR4 spectroscopic sample matching our emission line criteria of
       5$\sigma$ detections of H$\alpha$,
       H$\beta$, [NII]6584, and 3$\sigma$ detections of [OIII]5007 and [OII]3727. \\  \\ 
 
     2. For our emission line sample of $\sim8,000$ {\sl GALEX}/SDSS star forming galaxies, 
    we have examined each galaxy's oxygen abundance for three strong line abundance measurements. These are calibrated off photo-ionization models, M91, 
   {\it $T_e$} determined abundances, O3N2,
   and the Bayesian likelihood estimates, T04. We compare each abundance method as a function of both M$_*$ and SFR/M$_*$. 
    Compared to the other two methods O3N2 is found to increasingly estimate lower oxygen abundances 
    for galaxies with higher SFR/M$_*$ since it is not corrected for ionization parameter.\\ \\   

    3. We investigate the relationship between N/O and O/H 
    using the three different O/H diagnostics and the strong line calculation of N/O from \citet{pag92}. We use the specific SFR  
    derived from SED fits to the $7-$band {\sl GALEX}$+$SDSS photometry to indicate of the strength of the starbursts in each galaxy
    over the last 100Myr.\\ \\  

    4. Star forming galaxies that are currently forming a
       large percentage of their stellar mass, as parametrized by SFR/M$_*$, have smaller  values of N/O at a given metallicity for 
       for all three metallicity diagnostics, supporting the results of \citet{contini}.  
       This trend spans the metallicity range of $\sim0.6$ dex from $12+log O/H\sim8.4$ to  $12+log O/H \sim9.0$~dex and suggests the scenario
        that the scatter in N/O ratio for galaxies of similar metallicities is due to the ratio of current to past averaged SFR.
        The observed effect is modest, since the change in N/O is of the order of the abundance uncertainties. However the abundance dispersion could also arise
    due to varying mixing times of the newly synthesized oxygen into the ISM.   More realistic  and consistent metallicity
        diagnostics are required to further test this result. \\

\acknowledgments

GALEX is a NASA Small Explorer, launched in April 2003. We
gratefully acknowledge NASA's support for construction,
operation, and science analysis for the GALEX mission,
developed in cooperation with the CNES of France and the
Korean Ministry of Science and Technology. Funding for the
creation and distribution of the SDSS Archive has been
provided by the Alfred P. Sloan Foundation, the Participating
Institutions, NASA, NSF, DoE, Monbukagakusho, and the Max
Planck Society.

{\it Facilities:} \facility{GALEX}

\clearpage

\begin{figure}
\plotone{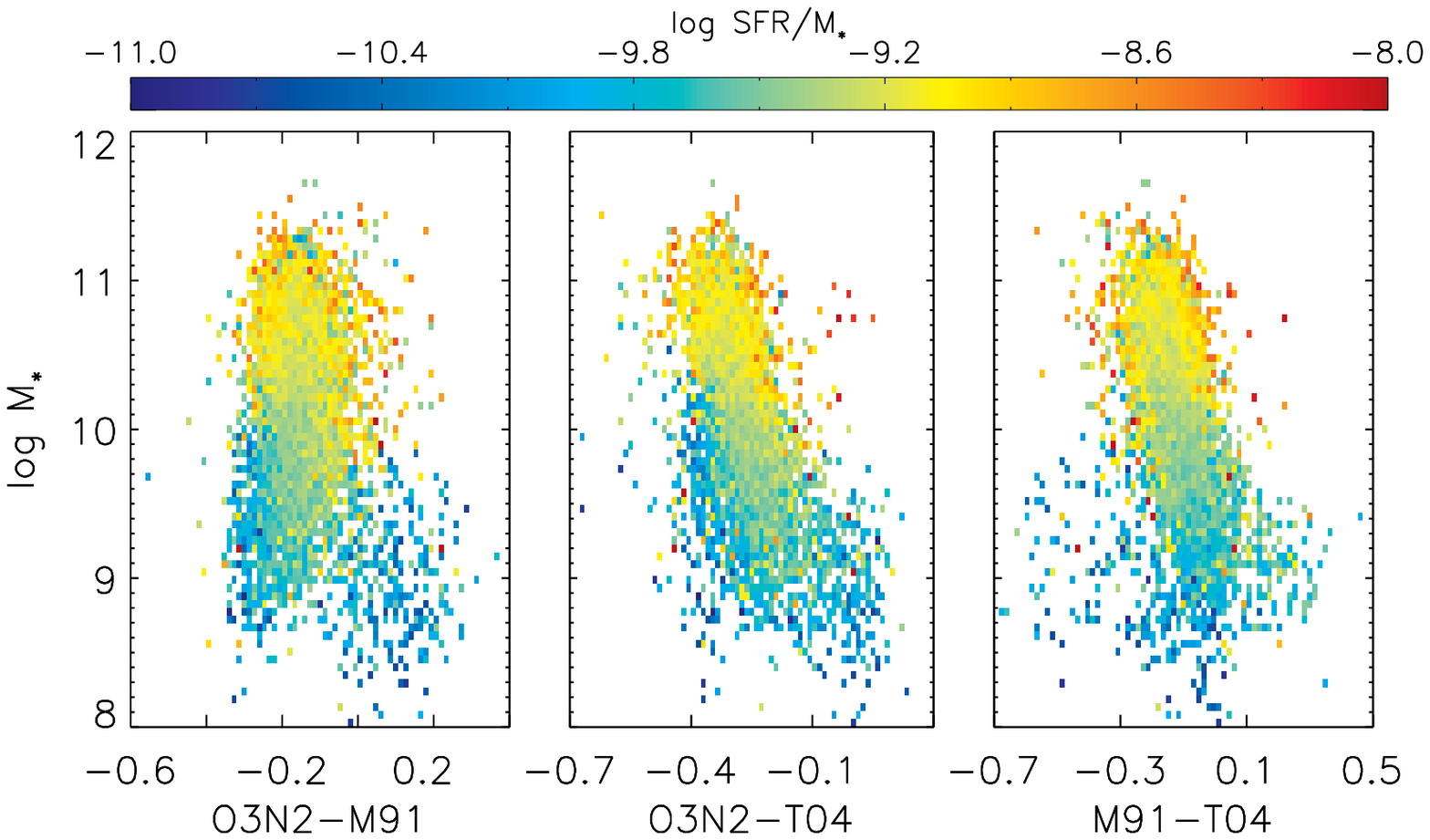}
\caption{Comparison of abundances from three different diagnostics: O3N2, M91, and T04 as a function of stellar mass.
        The data points are converted into a 75 by 75 pixel image. The mean specific SFR value of the points in each pixel is
        calculated and byte-scaled into true color. The difference between T04 and the other two diagnostics shows a dependence on mass
        since T04 estimates an increasingly larger metallicity at higher stellar masses. 
        }
\end{figure}


\begin{figure}
\plotone{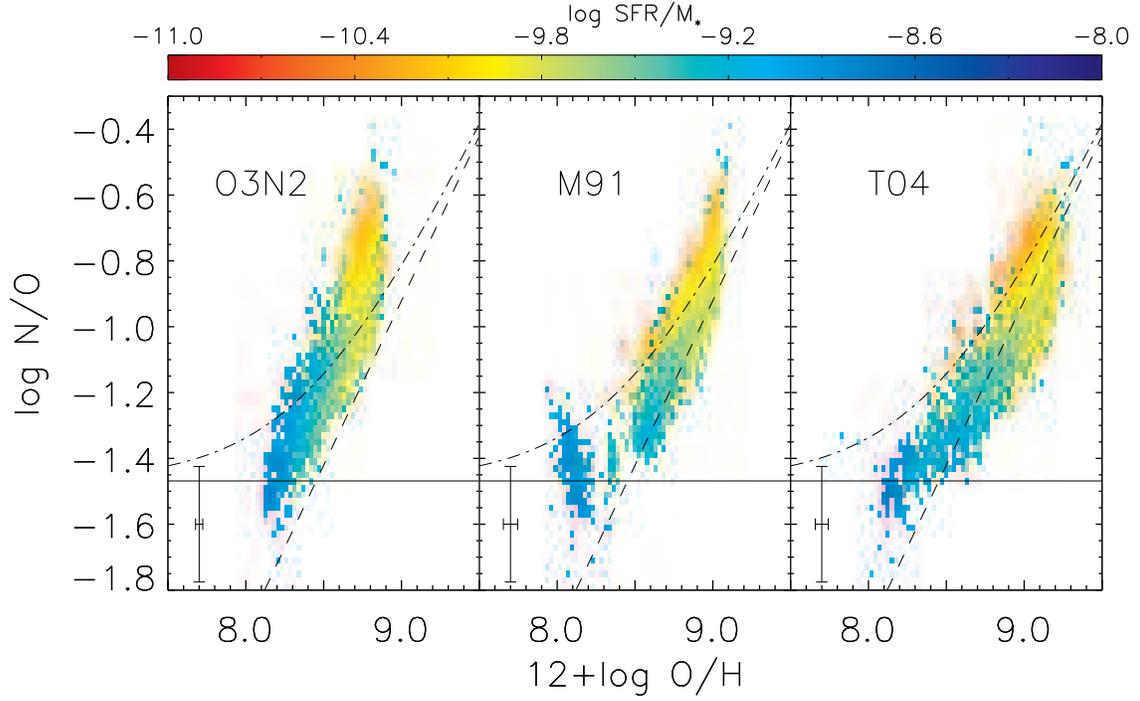}
\caption{N/O versus O/H for three different methods of  abundance determination. The points are binned into a 75 by 75 pixel image, with the 
          mean value of specific star formation rate (SFR/M$_*$) calculated from the points in each pixel and byte scaled into a true color representation.
          The specific star formation rate is an indicator of the star formation history of the galaxy.
          The plot shows the general trend that galaxies with similar metallicities have lower N/O ratios for larger values of SFR/M$_*$.
          This trend supports the time delay scenario where the the bulk of the oxygen is released from short lived massive stars, 
              and the release of the bulk of the nitrogen from longer lived intermediate mass stars.}
\end{figure}
\clearpage

\begin{figure}
\plotone{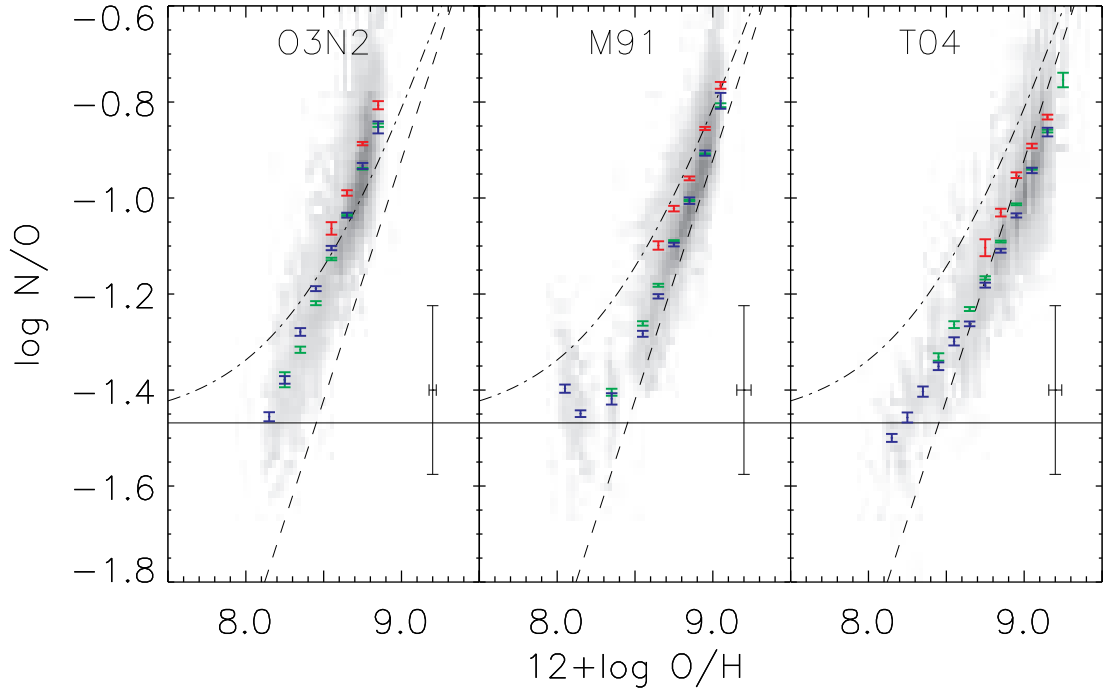}
\caption{2D Histogram of N/O versus O/H. The red points represent the mean N/O values for objects with SFR/M$_*< -10.1$ taken for metallicity increments of
          $0.1$ dex. The green points 
     represent objects with $ -10.1<$SFR/M$_*< -9.1$, and the blue data points represent galaxies with SFR/M$_*> -.9.1$. The abundance methods of O3N2, M91 and 
     T04 all show that the starbursts (having values of  SFR/M$_*> -10.1$) have lower values of N/O than galaxies of similar metallicity
     that are currently not forming as large a fraction of their stellar mass.}
\end{figure}
\clearpage

\begin{figure}
\plotone{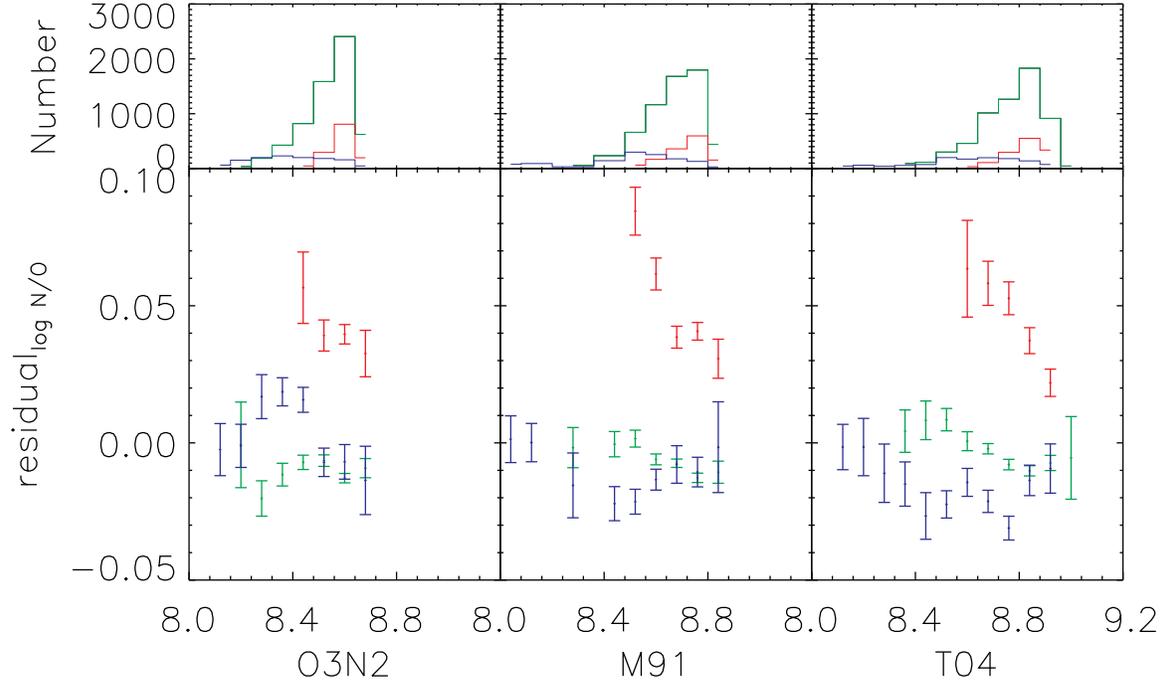}
\caption{Difference of  mean N/O ratios between three sub-samples of specific star formation rate and the mean N/O of the entire sub-sample taken
        for metallicity increments of $0.1$dex. 
         The red points represent the mean N/O values for objects with SFR/M$_*< -10.1$, the green points 
     represent objects with $-10.1<$SFR/M$_*< -9.1$, and the blue data points represent galaxies with SFR/M$_*> -9.1$. The abundance methods of O3N2, M91 and 
     T04 all show that galaxies with the lowest specific star formation rates  (having values of SFR/M$_* < -10.1$) have higher values of N/O than galaxies of 
      similar metallicity with  log SFR/M$_* > -10.1$. At intermediate metallicities the most extreme starbursts (log SFR/M$_* > -9.1$) on average have slightly 
       lower N/O  ratios than
      galaxies with average specific star formation rates ($-10.1<$log SFR/M$_*<-9.1$ by $0.02$ dex presumably due to oxygen released by the high mass stars formed
       in the starburst}.
\end{figure}
\clearpage

\begin{figure}
\plotone{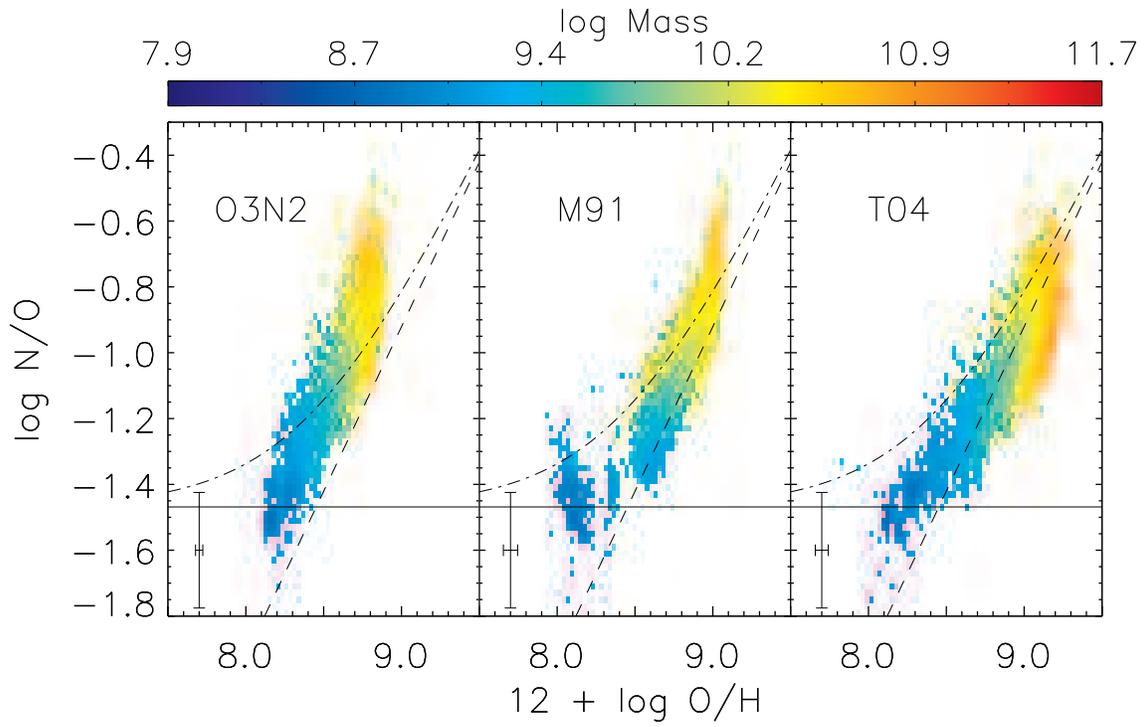}
\caption{N/O versus O/H scaled with Log M$_{*}$. This shows that  M$_{*}$ increases with increasing metallicity, and tends to have little dependence on N/O at a given
          metallicity.}
\end{figure}
\clearpage

\begin{figure}
\plotone{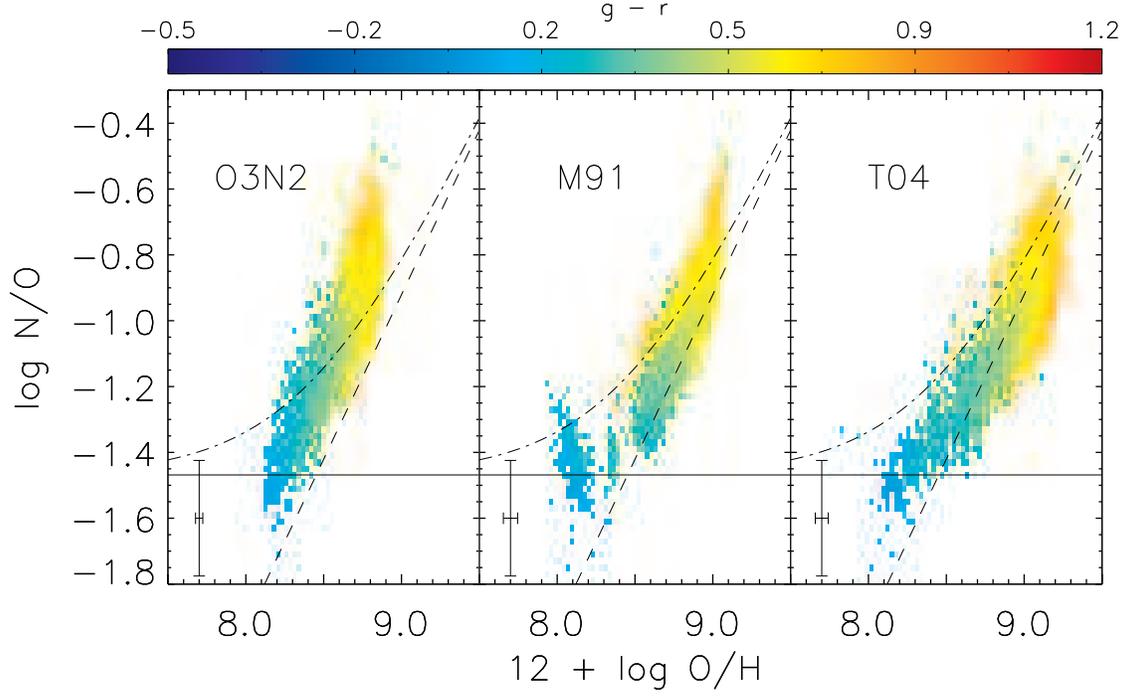}
\caption{N/O versus O/H, scaled with {\it g-r}. This plot shows that the {\it g-r} color of galaxies is redder for galaxies with higher metallicities. 
         Also for  galaxies with similar metallicities, those galaxies with larger values of N/O have a redder color. This is because redder galaxies have increasingly
         declining SFRs, where the intermediate mass stars from previous star formation events have released nitrogen into the ISM. This confirms the result of
         \citet{vzee06} who found a similar trend for dwarf galaxies. }
\end{figure}
\clearpage

\begin{figure}
\plotone{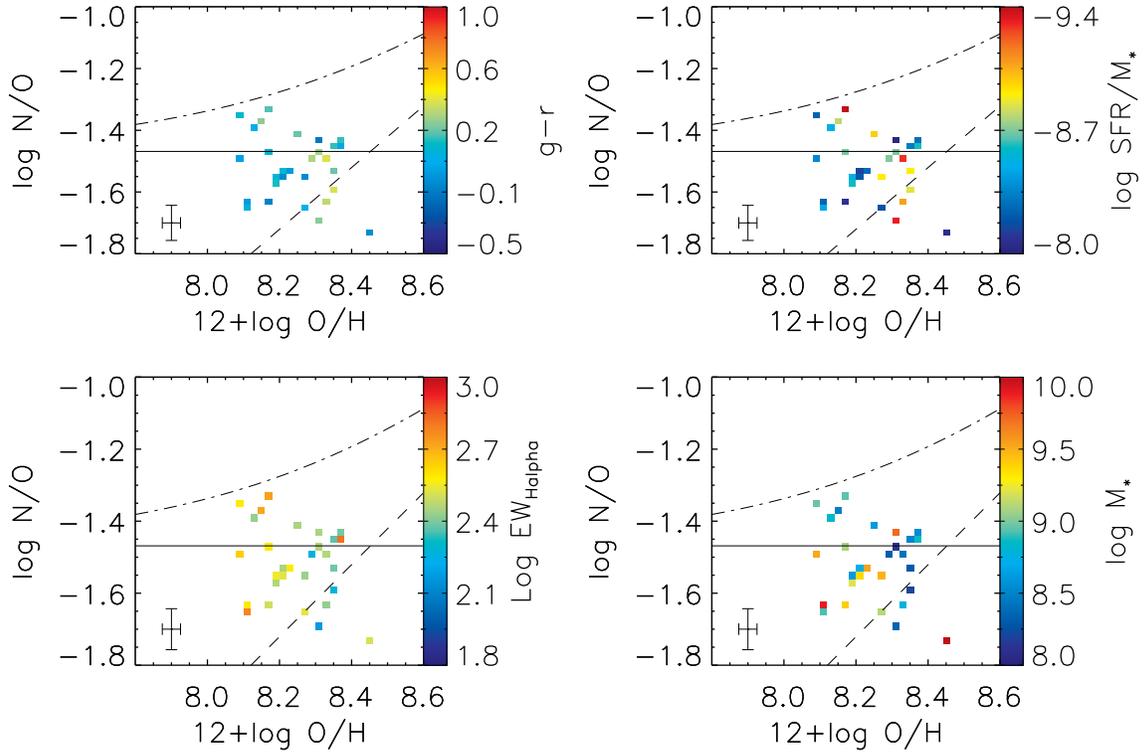}
\caption{N/O versus O/H  for abundances calculated using the {\it $T_e$} method. Overall the galaxies with $\log$ N/O$<-1.5$ on average tend to be slightly bluer, with 
         higher equivalent widths, and with specific star formation rates $0.12$ dex higher 
          than galaxies with $\log$ N/O$>-1.5$. The {\it $T_e$} oxygen abundances show little correlation with 
         stellar mass, but this is probably due to small sample size and errors in the derived abundances. A larger sample size of galaxies with abundances 
        derived from the  {\it $T_e$} method with a greater range of equivalent widths, stellar masses, and specific star formation rates, are needed to confirm
         the results from O3N2, M91, and T04.}
\end{figure}
\clearpage

\begin{deluxetable}{lll}
\tabletypesize{\scriptsize}
\tablecaption{Average Galaxy Parameter Errors}
\tablewidth{0pc}
\tablecolumns{3}
\tablehead{
\colhead{Parameter} &\colhead{$<value>$}  &\colhead{$<     \sigma>$}
}
\startdata
log M$_*$   &10.10  &0.075 \\
$<log SFR>_{100 Myr}$   &0.48 &0.20\\
$<\log$ SFR/M$_*>_{100 Myr}$  &-9.6 &0.19\\ 
log b                  &-0.24   &0.22\\
A$_{FUV}$             &2.40 &0.56\\
A$_{NUV}$               &1.78  &0.43\\
\enddata

\end{deluxetable}

\begin{deluxetable}{lrllrrlrl}
\tabletypesize{\scriptsize}
\tablecaption{${\it T_e}$ Abundances}
\tablewidth{0pc}
\tablecolumns{9}
\tablehead{ \colhead{$\alpha$J2000} &\colhead{dJ2000} &\colhead{$12+\log O/H$}  &\colhead{log N/O}
    &\colhead{log $SFR/M_*$}  &\colhead{$M_*$}   &\colhead{$F_{BURST}~ ^{1}$}  &\colhead{{\it g-r}} &\colhead{$\log EW_{H\alpha}$} 
    }
\startdata
20$^{h}$40$^{m}$18.047$^s$ & 01$^o$03$^{\prime}$24$^{\prime\prime}$.59 &8.31 $\pm$0.023 &-1.43 $\pm$0.047 &-8.19 $\pm$0.087 & 8.28 $\pm$0.057 &0.00$^{0.65}_{0.00}$ &-0.13 &2.355\\
20$^{h}$52$^{m}$51.797$^s$ & 00$^o$16$^{\prime}$26$^{\prime\prime}$.00 &8.16 $\pm$0.028 &-1.38 $\pm$0.057 &-8.89 $\pm$0.170 & 9.54 $\pm$0.115 &0.10$^{0.15}_{0.00}$ & 0.31 &2.091\\
21$^{h}$18$^{m}$29.846$^s$ & 00$^o$30$^{\prime}$59$^{\prime\prime}$.55 &8.18 $\pm$0.017 &-1.56 $\pm$0.049 &-9.16 $\pm$0.178 & 9.26 $\pm$0.115 &0.00$^{0.15}_{0.00}$ & 0.39 &2.506\\
22$^{h}$07$^{m}$07.888$^s$ & 00$^o$46$^{\prime}$58$^{\prime\prime}$.78 &8.10 $\pm$0.030 &-1.62 $\pm$0.086 &-8.39 $\pm$0.150 & 8.11 $\pm$0.082 &0.00$^{0.51}_{0.00}$ & 0.00 &2.216\\
22$^{h}$12$^{m}$23.328$^s$ & 00$^o$03$^{\prime}$39$^{\prime\prime}$.86 &8.24 $\pm$0.031 &-1.40 $\pm$0.062 &-9.06 $\pm$0.247 & 9.41 $\pm$0.128 &0.05$^{0.32}_{0.00}$ & 0.26 &2.354\\
21$^{h}$50$^{m}$29.868$^s$ & 00$^o$32$^{\prime}$01$^{\prime\prime}$.26 &8.24 $\pm$0.032 &-1.53 $\pm$0.070 &-8.24 $\pm$0.173 & 8.19 $\pm$0.105 &0.00$^{0.56}_{0.00}$ &-0.09 &2.349\\
20$^{h}$46$^{m}$56.140$^s$ & 00$^o$50$^{\prime}$37$^{\prime\prime}$.63 &8.27 $\pm$0.032 &-1.66 $\pm$0.070 &-8.34 $\pm$0.192 & 8.90 $\pm$0.140 &0.38$^{0.73}_{0.00}$ & 0.10 &2.279\\
10$^{h}$53$^{m}$42.546$^s$ & 00$^o$09$^{\prime}$45$^{\prime\prime}$.13 &8.19 $\pm$0.024 &-1.55 $\pm$0.053 &-8.66 $\pm$0.188 & 9.41 $\pm$0.113 &0.18$^{0.47}_{0.07}$ & 0.14 &2.258\\
17$^{h}$09$^{m}$22.632$^s$ & 61$^o$48$^{\prime}$51$^{\prime\prime}$.25 &8.36 $\pm$0.026 &-1.45 $\pm$0.048 &-8.47 $\pm$0.155 & 9.52 $\pm$0.118 &0.25$^{0.63}_{0.15}$ & 0.22 &2.440\\
00$^{h}$53$^{m}$00.523$^s$ & 15$^o$01$^{\prime}$29$^{\prime\prime}$.73 &8.19 $\pm$0.021 &-1.58 $\pm$0.047 &-8.28 $\pm$0.153 & 8.36 $\pm$0.062 &0.00$^{0.77}_{0.00}$ & 0.00 &2.183\\
08$^{h}$01$^{m}$43.632$^s$ & 44$^o$54$^{\prime}$58$^{\prime\prime}$.41 &8.37 $\pm$0.025 &-1.46 $\pm$0.052 &-8.73 $\pm$0.092 & 9.16 $\pm$0.085 &0.10$^{0.24}_{0.00}$ & 0.04 &1.985\\
08$^{h}$20$^{m}$01.714$^s$ & 50$^o$50$^{\prime}$39$^{\prime\prime}$.20 &8.35 $\pm$0.020 &-1.59 $\pm$0.045 &-8.94 $\pm$0.120 & 9.82 $\pm$0.053 &0.00$^{0.14}_{0.00}$ & 0.43 &2.537\\
08$^{h}$47$^{m}$03.007$^s$ & 54$^o$50$^{\prime}$39$^{\prime\prime}$.45 &8.22 $\pm$0.019 &-1.53 $\pm$0.060 &-8.30 $\pm$0.162 & 9.28 $\pm$0.125 &0.41$^{0.43}_{0.09}$ & 0.15 &2.356\\
03$^{h}$05$^{m}$39.705$^s$ &-08$^o$39$^{\prime}$05$^{\prime\prime}$.24 &8.20 $\pm$0.027 &-1.55 $\pm$0.083 &-8.36 $\pm$0.163 & 8.65 $\pm$0.092 &0.10$^{0.51}_{0.00}$ &-0.06 &2.279\\
12$^{h}$05$^{m}$14.725$^s$ & 66$^o$16$^{\prime}$57$^{\prime\prime}$.80 &8.32 $\pm$0.030 &-1.50 $\pm$0.058 &-9.34 $\pm$0.072 & 9.61 $\pm$0.047 &0.00$^{0.03}_{0.00}$ & 0.44 &2.346\\
10$^{h}$23$^{m}$19.567$^s$ & 02$^o$49$^{\prime}$41$^{\prime\prime}$.53 &8.08 $\pm$0.031 &-1.35 $\pm$0.059 &-8.39 $\pm$0.243 & 9.05 $\pm$0.140 &0.05$^{0.31}_{0.00}$ & 0.16 &2.218\\
11$^{h}$36$^{m}$55.796$^s$ & 03$^o$33$^{\prime}$33$^{\prime\prime}$.40 &8.34 $\pm$0.024 &-1.54 $\pm$0.047 &-9.00 $\pm$0.158 & 9.73 $\pm$0.117 &0.06$^{0.19}_{0.00}$ & 0.26 &2.422\\
08$^{h}$39$^{m}$14.949$^s$ & 48$^o$15$^{\prime}$18$^{\prime\prime}$.24 &8.17 $\pm$0.027 &-1.47 $\pm$0.053 &-8.51 $\pm$0.217 & 8.49 $\pm$0.112 &0.00$^{0.37}_{0.00}$ & 0.13 &2.136\\
09$^{h}$46$^{m}$30.590$^s$ & 55$^o$35$^{\prime}$41$^{\prime\prime}$.81 &8.23 $\pm$0.031 &-1.53 $\pm$0.062 &-8.58 $\pm$0.140 & 8.76 $\pm$0.075 &0.00$^{0.21}_{0.00}$ & 0.08 &2.084\\
14$^{h}$05$^{m}$01.154$^s$ & 04$^o$31$^{\prime}$26$^{\prime\prime}$.13 &8.46 $\pm$0.027 &-1.74 $\pm$0.054 &-8.23 $\pm$0.105 & 8.00 $\pm$0.035 &0.00$^{0.65}_{0.00}$ &-0.06 &2.297\\
14$^{h}$36$^{m}$48.204$^s$ & 04$^o$02$^{\prime}$59$^{\prime\prime}$.92 &8.17 $\pm$0.031 &-1.34 $\pm$0.060 &-9.44 $\pm$0.070 & 9.06 $\pm$0.082 &0.00$^{0.02}_{0.00}$ & 0.23 &2.077\\
14$^{h}$46$^{m}$10.316$^s$ & 03$^o$39$^{\prime}$21$^{\prime\prime}$.55 &8.31 $\pm$0.016 &-1.46 $\pm$0.032 &-8.71 $\pm$0.162 & 9.75 $\pm$0.100 &0.16$^{0.26}_{0.06}$ & 0.28 &2.437\\
14$^{h}$54$^{m}$24.609$^s$ & 03$^o$59$^{\prime}$25$^{\prime\prime}$.20 &8.29 $\pm$0.020 &-1.49 $\pm$0.041 &-8.81 $\pm$0.185 & 9.71 $\pm$0.075 &0.00$^{0.47}_{0.00}$ & 0.36 &2.543\\
00$^{h}$52$^{m}$49.794$^s$ &-08$^o$41$^{\prime}$33$^{\prime\prime}$.93 &8.10 $\pm$0.034 &-1.49 $\pm$0.070 &-8.51 $\pm$0.202 & 8.48 $\pm$0.145 &0.00$^{0.46}_{0.00}$ & 0.02 &2.163\\
01$^{h}$38$^{m}$44.917$^s$ &-08$^o$35$^{\prime}$40$^{\prime\prime}$.69 &8.18 $\pm$0.017 &-1.62 $\pm$0.045 &-8.24 $\pm$0.235 & 8.60 $\pm$0.033 &0.42$^{0.68}_{0.00}$ &-0.09 &2.303\\
01$^{h}$47$^{m}$21.680$^s$ &-09$^o$16$^{\prime}$46$^{\prime\prime}$.23 &8.31 $\pm$0.018 &-1.70 $\pm$0.050 &-9.37 $\pm$0.145 & 9.68 $\pm$0.115 &0.03$^{0.12}_{0.02}$ & 0.30 &2.642\\
02$^{h}$03$^{m}$56.913$^s$ &-08$^o$07$^{\prime}$58$^{\prime\prime}$.48 &8.37 $\pm$0.018 &-1.43 $\pm$0.036 &-8.43 $\pm$0.162 & 9.48 $\pm$0.090 &0.25$^{0.52}_{0.00}$ & 0.21 &2.414\\
22$^{h}$58$^{m}$33.743$^s$ & 00$^o$56$^{\prime}$30$^{\prime\prime}$.53 &8.13 $\pm$0.032 &-1.39 $\pm$0.081 &-8.65 $\pm$0.270 & 9.17 $\pm$0.193 &0.17$^{0.55}_{0.00}$ & 0.08 &2.377\\
23$^{h}$29$^{m}$32.117$^s$ & 00$^o$34$^{\prime}$26$^{\prime\prime}$.91 &8.32 $\pm$0.029 &-1.63 $\pm$0.079 &-9.15 $\pm$0.065 & 9.23 $\pm$0.100 &0.00$^{0.06}_{0.00}$ & 0.39 &2.382\\
22$^{h}$53$^{m}$56.829$^s$ & 10$^o$13$^{\prime}$00$^{\prime\prime}$.29 &8.31 $\pm$0.031 &-1.46 $\pm$0.060 &-8.88 $\pm$0.092 &10.04 $\pm$0.108 &0.10$^{0.10}_{0.01}$ & 0.41 &2.252\\
10$^{h}$21$^{m}$32.505$^s$ & 61$^o$44$^{\prime}$04$^{\prime\prime}$.52 &8.16 $\pm$0.023 &-1.48 $\pm$0.056 &-9.08 $\pm$0.207 & 9.36 $\pm$0.123 &0.06$^{0.22}_{0.00}$ & 0.16 &2.294\\
08$^{h}$20$^{m}$10.558$^s$ & 37$^o$43$^{\prime}$54$^{\prime\prime}$.34 &8.11 $\pm$0.026 &-1.66 $\pm$0.060 &-8.60 $\pm$0.162 & 9.09 $\pm$0.077 &0.00$^{0.38}_{0.00}$ & 0.13 &2.022\\
21$^{h}$19$^{m}$58.308$^s$ & 00$^o$52$^{\prime}$33$^{\prime\prime}$.52 &8.26 $\pm$0.016 &-1.55 $\pm$0.046 &-9.00 $\pm$0.435 & 8.52 $\pm$0.015 &0.07$^{0.07}_{0.00}$ &-0.06 &2.375
\enddata
\tablenotetext{1}{The superscripts for $F_{BURST}$ (Fraction of stellar mass formed in starbursts over the last 100Myr) list the 97.5 percentile values, and 
       the suscripts list the 2.5 percentile values.}
\end{deluxetable}

\begin{deluxetable}{llrlrrr}
\tabletypesize{\scriptsize}
\tablecaption{O3N2 mean N/O}
\tablewidth{0pc}
\tablecolumns{7}
\tablehead{ \colhead{$12+\log O/H$} &\colhead{Sub-sample$^1$} &\colhead{\#of galaxies} 
    &\colhead{$\log N/O_{mean}$} &\colhead{$\log SFR/M_{_*mean}$}  &\colhead{$\log M_{_*mean}$}
    &\colhead{${\it g-r}_{mean}$}
    }
\startdata
    &  a&   47 & -1.06 $\pm$0.013 &-10.49 $\pm$0.055 &10.14 $\pm$0.081 &0.69 $\pm$0.023\\
8.55&  b&  873 & -1.13 $\pm$0.003 & -9.49 $\pm$0.008 & 9.74 $\pm$0.011 &0.43 $\pm$0.003\\
    &  c&  149 & -1.10 $\pm$0.005 & -8.88 $\pm$0.018 & 9.88 $\pm$0.029 &0.36 $\pm$0.007\\ \\

    &  a&  295 & -0.99 $\pm$0.006 &-10.34 $\pm$0.014 &10.27 $\pm$0.025 &0.71 $\pm$0.006\\
8.65&  b& 1638 & -1.04 $\pm$0.002 & -9.62 $\pm$0.006 &10.05 $\pm$0.009 &0.52 $\pm$0.002\\
    &  c&  132 & -1.04 $\pm$0.006 & -8.89 $\pm$0.020 &10.06 $\pm$0.028 &0.43 $\pm$0.008\\\\

    &  a&  809 & -0.89 $\pm$0.004 &-10.33 $\pm$0.008 &10.53 $\pm$0.013 &0.74 $\pm$0.004\\
8.75&  b& 2459 & -0.94 $\pm$0.002 & -9.71 $\pm$0.005 &10.38 $\pm$0.007 &0.60 $\pm$0.002\\
    &  c&  109 & -0.93 $\pm$0.008 & -8.88 $\pm$0.024 &10.48 $\pm$0.034 &0.55 $\pm$0.011\\\\

    &  a&  197 & -0.81 $\pm$0.008 &-10.34 $\pm$0.021 &10.58 $\pm$0.023 &0.74 $\pm$0.007\\
8.85&  b&  637 & -0.85 $\pm$0.003 & -9.68 $\pm$0.009 &10.50 $\pm$0.014 &0.63 $\pm$0.004\\
    &  c&   36 & -0.85 $\pm$0.015 & -8.92 $\pm$0.030 &10.48 $\pm$0.050 &0.59 $\pm$0.015\\
\enddata
  \tablenotetext{1}{Sub-sample a: galaxies with  SFR/M$_* <-10.1$,   Sub-sample b: galaxies with $-10.1<$SFR/M$_* < -9.1$, Sub-sample c galaxies with SFR/M$_*>$-9.1}
\end{deluxetable}

\begin{deluxetable}{llccccc}
\tabletypesize{\scriptsize}
\tablecaption{M91 mean N/O}
\tablewidth{0pc}
\tablecolumns{7}
\tablehead{ \colhead{12$+$log O/H} &\colhead{Sub-sample$^1$}  &\colhead{\#of galaxies} 
     &\colhead{log N/O$_{mean}$} &\colhead{log SFR/M$_{*mean}$}  &\colhead{log M$_{*mean}$}
   &\colhead{{\it g-r}$_{mean}$}
   }
\startdata
     &  a &   64&-1.10 $\pm$0.009 &-10.48 $\pm$0.044 &10.12 $\pm$0.085 &0.69 $\pm$0.024 \\
 8.65&  b &  733&-1.18 $\pm$0.003 & -9.47 $\pm$0.009 & 9.62 $\pm$0.017 &0.42 $\pm$0.004\\
     &  c &  229&-1.21 $\pm$0.005 & -8.85 $\pm$0.016 & 9.58 $\pm$0.030 &0.31 $\pm$0.007\\\\
    
     &  a &  169&-1.02 $\pm$0.006 &-10.40 $\pm$0.024 &10.27 $\pm$0.036 &0.71 $\pm$0.009\\
 8.75&  b & 1228&-1.09 $\pm$0.002 & -9.55 $\pm$0.007 & 9.92 $\pm$0.012 &0.48 $\pm$0.004\\
     &  c &  191&-1.10 $\pm$0.004 & -8.88 $\pm$0.016 & 9.92 $\pm$0.027 &0.37 $\pm$0.008\\\\

     &  a &  360&-0.96 $\pm$0.004 &-10.33 $\pm$0.011 &10.37 $\pm$0.023 &0.72 $\pm$0.005\\
 8.85&  b & 1733&-1.01 $\pm$0.001 & -9.64 $\pm$0.006 &10.17 $\pm$0.010 &0.55 $\pm$0.003\\
     &  c &  126&-1.00 $\pm$0.009 & -8.84 $\pm$0.024 &10.13 $\pm$0.041 &0.46 $\pm$0.010\\\\

     &  a &  597&-0.86 $\pm$0.003 &-10.33 $\pm$0.009 &10.55 $\pm$0.015 &0.74 $\pm$0.004\\
 8.95&  b & 1839&-0.91 $\pm$0.002 & -9.70 $\pm$0.006 &10.39 $\pm$0.009 &0.60 $\pm$0.002\\
     &  c &   89&-0.91 $\pm$0.007 & -8.89 $\pm$0.024 &10.38 $\pm$0.041 &0.53 $\pm$0.012\\
 \enddata
 \end{deluxetable}
\tablenotetext{1}{Sub-sample a: galaxies with  SFR/M$_* <-10.1$,   Sub-sample b: galaxies with $-10.1<$SFR/M$_* < -9.1$, Sub-sample c galaxies with SFR/M$_*>$-9.1}

 \begin{deluxetable}{llccccc}
 \tabletypesize{\scriptsize}
 \tablecaption{T04 mean N/O}
 \tablewidth{0pc}
 \tablecolumns{7}
 \tablehead{ \colhead{$12+\log O/H$} &\colhead{Sub-sample$^1$}  &\colhead{\#of galaxies} 
   &\colhead{$\log N/O_{mean}$} &\colhead{$\log M_{*mean}$}  &\colhead{$log L_{FUV}/M_{*mean}$}
     &\colhead{${\it g-r}_{mean}$}
     }
  \startdata
     &  a&   35 &-1.10 $\pm$0.018 &-10.38 $\pm$0.045 & 9.96 $\pm$0.102 &0.64 $\pm$0.028\\
 8.75&  b&  506 &-1.17 $\pm$0.003 & -9.47 $\pm$0.010 & 9.63 $\pm$0.014 &0.42 $\pm$0.004\\
     &  c&  128 &-1.18 $\pm$0.006 & -8.85 $\pm$0.021 & 9.67 $\pm$0.029 &0.32 $\pm$0.008\\\\

     &  a&  110 &-1.03 $\pm$0.008 &-10.41 $\pm$0.032 &10.03 $\pm$0.042 &0.67 $\pm$0.011 \\
 8.85&  b& 1060 &-1.09 $\pm$0.002 & -9.55 $\pm$0.007 & 9.84 $\pm$0.010 &0.47 $\pm$0.003\\
     &  c&  154 &-1.11 $\pm$0.005 & -8.86 $\pm$0.019 & 9.90 $\pm$0.027 &0.37 $\pm$0.008\\\\

     &  a&  297 &-0.95 $\pm$0.006 &-10.34 $\pm$0.013 &10.31 $\pm$0.022 &0.71 $\pm$0.006\\
 8.95&  b& 1313 &-1.01 $\pm$0.002 & -9.63 $\pm$0.007 &10.10 $\pm$0.009 &0.53 $\pm$0.003\\
     &  c&  133 &-1.04 $\pm$0.005 & -8.86 $\pm$0.022 &10.06 $\pm$0.033 &0.42 $\pm$0.009\\\\

     &  a&  551 &-0.89 $\pm$0.005 &-10.32 $\pm$0.009 &10.51 $\pm$0.015 &0.74 $\pm$0.004\\
 9.05&  b& 1874 &-0.94 $\pm$0.002 & -9.70 $\pm$0.006 &10.37 $\pm$0.008 &0.59 $\pm$0.002\\
     &  c&   95 &-0.95 $\pm$0.006 & -8.87 $\pm$0.027 &10.36 $\pm$0.036 &0.51 $\pm$0.011\\\\

     &  a&  338 &-0.83 $\pm$0.005 &-10.35 $\pm$0.015 &10.68 $\pm$0.017 &0.76 $\pm$0.005\\
 9.15&  b&  935 &-0.86 $\pm$0.003 & -9.71 $\pm$0.008 &10.60 $\pm$0.011 &0.65 $\pm$0.003\\
     &  c&   56 &-0.86 $\pm$0.011 & -8.93 $\pm$0.024 &10.59 $\pm$0.043 &0.60 $\pm$0.015\\
 \enddata
\tablenotetext{1}{Sub-sample a: galaxies with  SFR/M$_* <-10.1$,   Sub-sample b: galaxies with $-10.1<$SFR/M$_* < -9.1$, Sub-sample c galaxies with SFR/M$_*>$-9.1}
 \end{deluxetable}
\end{document}